\documentstyle[prl,aps,amsfonts,amssymb,twocolumn]{revtex}
\title{Breathing Modes and Hidden Symmetry of  Trapped 
Atoms in 2D} \author{ L.P. Pitaevskii$^{1,2,3}$ and A. Rosch$^1$}
\address{$^1$Institut f\"ur Theorie der Kondensierten Materie, 
Universit\"at Karlsruhe,
D--76128 Karlsruhe, Germany}
\address{$^2$Department of Physics, Technion, 32000 Haifa, Israel}
\address{$^3$Kapitza Institute for Physical Problems, 117334 Moscow, Russia}
\begin{document}
\draft

\twocolumn[\hsize\textwidth\columnwidth\hsize\csname 
@twocolumnfalse\endcsname %
\date{September 12, 1996} \maketitle

\begin{abstract}
Atoms confined in a harmonic potential show universal oscillations in 2D. 
We point out the connection of these ''breathing'' modes to the
presence of a hidden symmetry.
The underlying symmetry SO(2,1), i.e. the two dimensional
Lorentz group, allows pulsating solutions to be constructed 
for the interacting
quantum system and for the corresponding nonlinear Gross-Pitaevskii equation. 
 We point out how this symmetry can be used as a probe for
recently proposed experiments of trapped atoms in 2D.
\end{abstract}


\vskip1pc]
%

The problem of Bose-Einstein condensation in an external
 potential has received a lot of interest after the experimental observation 
 of the condensation in alkali-atom vapors
\cite{1}. In real experiments the trapping
 potential is approximately  harmonic with frequency $\omega_0$. This implies specific
 peculiarities in the behavior of the system. In recent papers
  \cite{pitaevskii2,benjamin,kagan}
 a few of these properties have been demonstrated. It has been
 shown \cite{pitaevskii2} that the nonlinear
 Gross-Pitaevskii (GP) equation  (\ref{1}) for a trapped 
 2D system possesses ``breathing`` oscillatory modes 
 with the  universal frequency $2\omega _0$, 
 describing a pulsation of the condensate. The same modes  show up in
 \cite{kagan}, where the authors were able to construct 
explicitly the
 time evolution of the GP equation  in a time dependent external potential
 in 2D.
 In \cite{benjamin} the authors discovered that 
 the energy spectrum of a system of trapped particles 
 interacting with a $1/r ^2$ potential is
 divided in sets of equidistant levels with the separation $2\omega _0$ again.
 To our understanding this interesting property has not yet been 
 properly explained and the connection between these different systems
 has not been established.

 In this paper we shall show that in these cases 
 the existence of these $2\omega _0$
 oscillations is ensured by a specific symmetry property of the system. 
 Proper use of this symmetry leads to a transformation which permits 
a set of  breathing mode wave functions to be
constructed algebraically
 not only for the $1/r ^2$ problem, but also for a local interaction
 in two dimensions.
Actually this symmetry is not only a property of the mean field theory
 as found in \cite{pitaevskii2,kagan} but of the full quantum theory.
We will show, that the oscillating solutions of \cite{kagan} are
  a
 continuous representation of the underlying group $SO(2,1)$.

 To understand the role of symmetry in the problem it is useful to
 consider as an instructive example a system of classical particles moving in
 a harmonic external potential
 $V_{\text{pot}} = \sum _i\frac {1}{2} m\omega_0 ^ 2 r_i^2$
 and interacting with a potential $V(\bbox{r}_i)$ 
 with the scaling property:
$V(\lambda \bbox{r}_i) = \lambda ^{n}V(\bbox{r}_i)$.

  Let us consider the quantity $I=\sum _i r_i^2$ such that $\partial _t I = 
  2\sum _i\bbox{r}_i\bbox{p}_i/m$. Following the usual derivation of the
  virial theorem in classical mechanics we get:
\begin{eqnarray}
\partial _t \sum _i\bbox{p}_i\bbox{r}_i&=& 
\sum _i(\partial _t\bbox{r}_i)\bbox{p}_i
-\sum _i\bbox{r}_i
\nabla _i(V+V_{\text{pot}}) \nonumber \\ &=&2T-nV-2V_{\text{pot}}.
\label{01}
\end{eqnarray}
We now see that for a potential with the scaling exponent $n=-2$ the right
hand side of (\ref{01}) takes the form $2E-2 m \omega_0^2 I$,
where $E$ is the total energy of the system. In this case one gets a closed
equation for $I$:
\begin{eqnarray}
\partial _t^2I=-4\omega_0 ^2I + 4E/m
\label{02}
\end{eqnarray}
with the obvious solution:
$I=A \cos (2\omega _0 t + \gamma ) + E/(m \omega_0^2).$
Thus the existence of the ``$2\omega _0$`` modes 
is connected with the $n=-2$ scaling of the interaction potential. 
In a 3D system the only potential possessing
this property is the $1/r^2$ interaction used in \cite{benjamin}.
 But in the quantum
$2D$ case the Fermi ``pseudo-potential`` 
$\frac{1}{2} g \delta^2 (\bbox{r})$ as used in the GP equation
  \cite{pitaevskii2,kagan}
gives the same scaling. 

The equations above can be rewritten by introducing (the notations
will become obvious later) $L^+=\sum _i\bbox{p}_i\bbox{r}_i /2+i(E-m \omega_0^2 I)/(2
\omega_0) $. We simply get $\partial_t L^+=i 2
\omega_0 L^+$. The phase of $L^+$ varies linearly with time:
\begin{eqnarray*}
\Phi(\bbox{p}_i,\bbox{r}_i) &\equiv& \frac{1}{2 \omega_0} \text{Im} \log
L^+ \\
&=&\frac{1}{2 \omega_0} \arctan \frac{(E-m \omega_0^2 I)/(2
\omega_0)}{\sum _i\bbox{p}_i\bbox{r}_i /2} \\
\Phi(\bbox{p}_i,\bbox{r}_i)&-&\Phi(\bbox{p}_i^0,\bbox{r}_i^0)=t-t_0,
\end{eqnarray*}
where $\bbox{p}_i^0,\bbox{r}_i^0$ are the coordinates at
$t=t_0$. $\Phi$ can now be used to determine the
``abbreviated action'' $S(E,\bbox{r}_i)$
which is a
function of the energy and the coordinates at the end of a path. It is
determined by the Hamilton-Jacobi equation $H(\partial_{\bbox{r}_i}
S,\bbox{r}_i)=E$ with $t=\partial_E S=\Phi(\partial_{\bbox{r}_i} S,\bbox{r})$.
In hyper-spherical coordinates in the space of all $\bbox{r}_i$ with
$r=\sqrt{I}$,
 $\Phi$ is only a function of $r$ and $\partial_r S$. Therefore we have
$S(E,\bbox{r}_i)=S(E,r)+S_0(\alpha_i)$
where the $\alpha_i$ denote all other coordinates, clearly showing
that the coordinate $r$ or $I$ totally separates. 


One of the most powerful methods in physics is the use of symmetries
and groups. One way is to use the invariance of the Hamilton or the
action under certain transformations, another is recognizing that the
Hamiltonian is a part of some larger algebra.  The most famous textbook
example is the algebraic solution of the harmonic oscillator using the
spectrum generating Heisenberg algebra $[H,a^{\pm}]=\pm \omega_0 a^\pm
$ (we put $\hbar=1$ throughout the paper). 

We will now discuss such a spectrum generating symmetry for the
(now quantum-mechanical) problem of interacting particles in a harmonic trap. First we will
consider the effect of a scaling-transformations for the Hamiltonian $
H_0=\sum_i -\frac{1}{2 m} \Delta_i+\sum_{i<j}
V(\bbox{r}_i-\bbox{r}_j)$ without an external potential:
\begin{eqnarray}
\bbox{r} &\rightarrow& \lambda \bbox{r},  
\Psi(\bbox{r})  \rightarrow   \lambda ^{d/2} 
\Psi( \lambda \bbox{r}) \nonumber \\
 H_0 &\rightarrow&  \frac{1}{\lambda ^2}
 \sum_i -\frac{1}{2 m} \Delta_i+
\sum_{i<j} V(\lambda (\bbox{r}_i-\bbox{r}_j)). 
\label{scaling}
\end{eqnarray}
$H_0$ is scale invariant if $V(\lambda \bbox{r})=
V(\bbox{r})/\lambda^2$. This is the case for an interaction of the form
$V(\bbox{r})=g/r^2$ in any dimension but as mentioned above also for
\begin{equation}
V(\bbox{r}-\bbox{r}')=\frac{1}{2} g \delta^2 (\bbox{r}-\bbox{r}')
\end{equation}
in two dimensions.  The first
case is known as the Calogero-Sutherland model \cite{calogero} and
widely investigated. All our results apply for both cases as
we will only use symmetry properties connected to
scale transformations, but we concentrate on the problem of bosons
with a local interaction and only shortly comment on the relation to
known results for the Calogero-Sutherland model. Actually the local interaction
is the most important case, at least for neutral atoms at low energies, where the
range of interaction is small compared to all other scales.
 The statistics of the
particles does not play a role.

It is important to note that a $\delta$-function interaction is not
well defined in two dimensions due to logarithmic ultraviolet
divergences which are cut off by the finite range $a_0$ of the
interaction.  This length $a_0$ obviously breaks scale invariance of
$H_0$ and will therefore modify our results.  Nevertheless this effect
will be small as long as $a_0$ is smaller than any other scale in the
system. This is most clearly seen in the classical wave-limit 
(i.e. the GP-equation) 
which we will discuss later, where such a problem is absent.

Adding an external potential $H=H_0+H_{\text{pot}}$,
$H_{\text{pot}}=\sum_i \frac{1}{2} m \omega_0^2 r_i^2$ obviously
breaks scale invariance, as $H_{\text{pot}} \rightarrow \lambda^2
H_{\text{pot}}$ under a scale transformation. However, due to a
special property of the harmonic oscillator, still a powerful spectrum
generating symmetry exists.

The important step is to recognize that the commutator of the harmonic
potential with the Hamiltonian or the time derivative of $\sum_i r_i^2$
is proportional to the generator of scale transformations.
\begin{eqnarray}
  [H_{\text{pot}},H]&=&[\sum_j \frac{1}{2} m \omega_0^2 r_j^2,
                        \sum_i -\frac{1}{2 m} \bbox{\nabla}^2_i] \nonumber \\
       &=&\sum_i \frac{1}{2} \omega_0^2 (\bbox{\nabla}_i \bbox{r}_i+
         \bbox{r}_i \bbox{\nabla}_i)=
       i  \omega_0^2 Q \label{q} \\
  Q&=&\sum_i \frac{1}{2} (\bbox{p}_i \bbox{r}_i+ \bbox{r}_i \bbox{p}_i). \nonumber
\end{eqnarray}
$Q$ is the generator of scale transformations as it describes the
translation of the coordinates $\bbox{r}_i$ by an amount proportional
to $\bbox{r}_i$.

We can collect our results (\ref{scaling},\ref{q}) in the following
algebra:
\begin{eqnarray}
[Q,H_0] &=& 2 i H_0,\  [ Q,H_{\text{pot}} ]=
 - 2 i H_{\text{pot}},\  [H_{\text{pot}},H] =
           i  \omega_0^2 Q \nonumber
\end{eqnarray}
or using
\begin{eqnarray}
 L_1=\frac{1}{2 \omega _0} (H_0-H_{\text{pot}}),  
L_2=\frac{Q}{2} \nonumber \\
L_3=\frac{1}{2 \omega_0}
(H_0+H_{\text{pot}})=\frac{1}{2 \omega_0} H, 
\end{eqnarray}
we get the algebra:
\begin{eqnarray}
 [ L_1,L_2 ]=-i L_3 ,
[L_2,L_3]=i L_1 , [L_3,L_1] =i L_2. 
\end{eqnarray}
This is the well known algebra of $SU(1,1)$ or $SO(2,1)$, the two
dimensional Lorentz group. $L_1$ and $L_2$ are the generators of the
two ''boosts'' and $L_3=\frac{1}{2 \omega_0} H$, i.e. the generator
of time-translations, is the analog of the generator of the rotation.
With $L^\pm = \frac{1}{\sqrt{2}}(L_1 \pm i L_2) $ this reads
\begin{eqnarray}
 [H,L^\pm]= \pm 2 \omega_0 L^\pm ,
 [L^+,L^-]= -  \frac{1}{2 \omega_0} H.
 \label{algebra}
\end{eqnarray}
Note the minus-sign in the last equation, indicating that the group is the
Lorentz group $SO(2,1)$ (or $SU(1,1)$) and not $SO(3)$.

One important consequence of this spectrum generating symmetry is
that the Hilbert-space will separate into irreducible representations
of the group. If the energy is bounded from below these are discrete
infinite-dimensional representations with no upper
bound. Starting from the lowest eigenstate in one of
the representations with energy $E_0$, 
$H |\Psi_0\rangle = E_0 |\Psi_0\rangle$ one
can construct higher states with energies $E_0 + n   2 \omega_0$,
$n=1,2,\dots$ by applying $L^+$ (use $H L^+ |\Psi_0\rangle=(L^+ H +2
\omega_0 L^+)|\Psi_0\rangle=(E_0+2 \omega_0) L^+|\Psi_0\rangle$). 
$|\Psi_0\rangle$ is
annihilated by $L^-$.  Obviously an infinite number of excitations
with energies $n 2 \omega_0$ exists which we will identify with the
breathing modes of the system.

Also the time dependence of all the operators, which are part of the
algebra can be given explicitly:
\begin{eqnarray}
 L^\pm(t)=e^{\mp i 2 w_0 t} L^\pm.
\end{eqnarray}
Defining the operator for the mean square displacement of the particles
$\hat{I}(t)=\sum \bbox{r}_i^2(t)$ one finds
\begin{eqnarray}
\hat{I}(t)&=&\frac{2}{ m \omega_0^2} H_{\text{pot}}(t) \nonumber \\
&=&\frac{1}{m \omega_0^2} \left(H-2  \omega_0 \sqrt{2} 
 \text{Re} \left( L^+ e^{- 2 w_0 t} \right)\right). \label{Iop}
\end{eqnarray}
This equation or the corresponding differential equation $\partial^2_t
{\hat{I}}=-(2 \omega_0)^2 (\hat{I}- H/(m \omega_0^2))$,
coinciding with the classical Eq. (\ref{02}),
 clearly show that
the variable $\hat{I}$ is separated from all other variables, in analogy
to the center of mass motion. This separation is known in the case of
the $1/r^2$ interaction \cite{benjamin}. As the frequency $\omega_0$
of the external potential obviously couples only to this coordinate
also the full nonlinear response to a change of the potential
can be calculated \cite{benjamin,kagan}.

For the expectation value we directly get
\begin{eqnarray}
  \label{I}
  I(t)=\langle\hat{I}(t)\rangle=I_0+A \cos(2 \omega_0 t+\gamma)
\end{eqnarray}
with $I_0=\langle H\rangle\!/(m \omega_0^2)$ and $A e^{-i
\gamma}=\sqrt{2}/(m \omega_0)
\langle L^+\rangle_{t=0}$. The same solution was found recently for the GP
equation in $D=2$ \cite{pitaevskii2,kagan}, but here we show
that it also holds for the full quantum system. 
It is valid not only for the expectation value
but directly for the operators (\ref{Iop}) and is due to a simple
underlying symmetry. We will discuss the change of this equation under
a group transformation later.

The Casimir operator of this group
$
C=H^2-(2  \omega_0)^2 (L_1^2+L_2^2)
$
commutes not only with the Hamiltonian, giving a new conserved quantity,
but also with the generators of the group and is the analog  of
the invariant line element in special relativity.

The classical wave limit of interacting bosons, 
i.e. the non-linear GP equation \cite{gross}, has recently
attracted considerable attention as it accurately describes the
Bose-condensate of trapped atoms and allows for reliable
analytic and especially numerical calculations \cite{stringari}. It is also an
interesting problem on its own for mathematical physics. We
will show that the previously discovered analytic
solutions  \cite{kagan} actually form a  continuous 
representation of the group
$SO(2,1)$ for the GP-equation.

In \cite{kagan} the following transformation of the solution of the
GP equation ($m=\hbar=1$)
\begin{eqnarray}
i\frac{\partial \Psi }{\partial t}  = - \frac{1}{2 }\Delta \Psi \!+\!
\frac{1}{2}  \omega_0 ^2 r^2\Psi  \!+\! 
g | \Psi  |^2 \Psi   \!-\! \mu \Psi .
\label{1}
\end{eqnarray}
was considered (in our notation):
\begin{eqnarray}
\Psi _2 (\bbox{r} , t) &=&\exp [i(b+cr^2 )]\frac{1}{\sqrt a}
\Psi _1 (\bbox{u},\tau),
 \label{2} \\
\bbox{u}&=&\frac{\bbox{r}}{\sqrt{a(t)}}, \tau(t) = \int^t \frac{dt'}{a(t')}
\label{2a}
\end{eqnarray}
If the  functions $a(t)$, $b(t)$ and $c(t)$ fulfill ($a_t=\partial_t a$ etc.):
\begin{eqnarray}
&&a_t=4ac, b_t=\mu \left (1-\frac{1}{a}\right )
\nonumber \\
&&
a_{tt} a-\frac{a_t^2}{2}+\frac{(2\tilde{\omega}_0)^2 }{2} a^2-\frac{(2\omega_0)^2 }{2}=0,
\label{7}
\end{eqnarray}
and if $\Psi_1$ is a solution of (\ref{1}) then $\Psi_2$ is also a solution of
 the GP-equation with a possibly
time dependent
frequency $\tilde{\omega}_0(t)$.
Differentiating (\ref{7}) gives the linear
 equation $a_{ttt}+(2 \tilde{\omega}_0)^2 a_t=
-4 \tilde{\omega}_0 \tilde{\omega}_{0t} a$ which once again demonstrates
the universal nature of these modes; the initial values must
 fulfill (\ref{7}).

We will now consider the case  $\tilde{\omega}_0 =\omega_0=const$
where the differential equations (\ref{7}) can 
can be solved directly: 
\begin{eqnarray} a\!&=&\!\sinh \eta \cos(2\omega_0 t+\gamma)+\cosh \eta.
\label{8}\\
\tau\!&=&\!\frac{1}{\omega_0} \left( \arctan \left(e^{-\eta} 
\tan\left(\omega_0 t\right) \right) +  \pi n\right)
 \nonumber \\
&=&\!\frac{1}{2 \omega_0} \Big(\arccos\!\left(
      \frac{\sinh \eta+\cosh \eta \cos 2  \omega_0 t}{\cosh \eta+
\sinh \eta \cos 2  \omega_0 t}\right)
    \text{sign}(\sin 2  \omega_0 t) \nonumber \\
&&\! \hphantom{\frac{1}{2 \omega_0} \Big(}+ 2 \pi n\Big) 
\label{9}
\end{eqnarray}
and $c(\eta,t),b(\eta,t)$ accordingly. 
The integer $n$ has to be chosen to get a continuous solution, 
$n(t)=[\omega_0 t/\pi+1/2]$, where $[x]$ denotes 
the greatest integer less than $x$.
A particularly important example is the case when the initial solution
is a static one, e.g. the groundstate, $\Psi _1=\Psi _0(\bbox{r})$. 
The transformation
builds then from such a static solution an oscillating breathing
solution.  In this case the parameter $\eta$ of the transformation
defines the relation of the energy of the new solution $E$ to the
static one $E_0$ according to $E=\cosh(\eta)E_0$.  In the following we
set the phase $\gamma=0$ in (\ref{8})  as a finite $\gamma$ can be
achieved by a simple translation in time.

Together with the time translations the discussed transformation form
a group, indeed a continuous representation of $SO(2,1)$. A general
group transformation $U_{t_2, \eta, t_1}$ can be described by a
initial translation in time by $t_1$, the above described scaling
transformation parameterized by $\eta$ and a second final translation
backwards in time by $t_2$.  This is the 
analog to the description of
rotations by Euler angles. With $\tau'=\tau(\eta,t-t_1)$ and
$a',b',c'$ accordingly we have
\begin{eqnarray}
  U_{t_2, \eta, t_1} \Psi =e^{i(b'+c' r^2 )}\frac{1}{\sqrt a'}
\Psi (\frac{1}{\sqrt{a'}} \bbox{r},\tau '+t_2).
\end{eqnarray}

It is now simple to work out the multiplication rules of this group -
they are the same as for two dimensional Lorentz transformations. For
example if one performs two successive ``Lorentz boosts'' in the same
direction one has to add the rapidities: 
\begin{equation}
U_{0,\eta_2,0} \circ
U_{0,\eta_1,0} =U_{0,\eta_1+\eta_2,0}.
\end{equation}
 We can see this by calculating
the corresponding function $a(t)$ for two transformations using
(\ref{9}):
\begin{eqnarray}
a&=&(\cosh \eta_2+\sinh \eta_2 \cos 2 \omega_0 t) \nonumber \\
&&\times \left(\cosh \eta_1
+\sinh \eta_1 \cos( 2 \omega_0 \tau(\eta_2,t))\right) \nonumber \\
 &=&\cosh(\eta_1+\eta_2)+\sinh(\eta_1+\eta_2) \cos 2 \omega_0 t.
\label{a}
\end{eqnarray}

To make the connection to the previously constructed algebra, one has
to calculate the generators of the group, i.e. the infinitesimal
transformations $\delta \eta \to 0$.
\begin{eqnarray}
  \label{10}
  a &\approx& 1+ \delta \eta \cos 2 \omega_0 t,\ \tau \approx t-
\delta \eta \frac{1}{2 \omega_0} \sin 2 \omega_0 t. \nonumber 
\end{eqnarray}
Under such a transformation the wave function changes by $\delta \Psi
=U_{0,\delta \eta,0}\Psi-\Psi$:
\begin{eqnarray}
\delta \Psi \approx -i \delta \eta \bigg[&&
    \frac{1}{4} (\bbox{r} \bbox{p}+ \bbox{p}\bbox{r}) 
    \cos 2 \omega_0 t  \nonumber \\
 && -   \frac{1}{2 \omega_0}
 \sin  2 \omega_0 t \left(i \frac{\partial}{\partial t}
        -2 \frac{1}{2} \omega_0^2 r^2 \right)\bigg] \Psi  
\end{eqnarray}
with $\bbox{p}=-i \bbox{\nabla}$.  This has to be compared with
$L_1=\frac{1}{4} (\bbox{r} \bbox{p}+ \bbox{p}\bbox{r})$ and
$L_2=\frac{1}{2 \omega_0} (H-2 H_{\text{pot}})$. Identifying $H$ with
$i \partial_t$ and noting that in the Heisenberg picture
$L_2(t)=\sqrt{2} \text{Im} L^+(t)=L_2 \cos 2 \omega_0 t-L_1 \sin 2
\omega_0 t$ we can identify $U_{0,\eta,0}$ with the transformation
generated by $L_2$. Accordingly $L_1$ generates the transformation
$U_{0,\eta,\pi/(2 \omega_0)}$ and $H$ obviously $U_{t,0,0}$.
Actually the solutions generated by $L_1$ and $L_2$ 
describe the breathing of the system, i.e. a pulsating motion.

To finish this section we directly
evaluate the effect of a group transformation to the mean square
displacement of the bosons $I(t)=\int r^2 |\Psi| d^2 r$. From
(\ref{9}) we know, that for any given solution $\Psi_0$, $I(t)$ is of
the form $I(t)=C (\cosh \eta_0 + \sinh \eta_0 \cos 2 \omega_0 t)$ as
long as the system does not collapse, i.e. as long as $I(t)>0$.
(Such a collapse is
 of course  possible only for an attractive interaction, i. e.
for $g<0$.)
 With
$\Psi'=U_{0,\eta,0} \Psi$ and using (\ref{a}) we get: 
\begin{eqnarray}
  I'(t)&=&\int r^2 |\Psi'(\bbox{r},t)| d^2 r= a(t) I\left(\tau(t)\right) \nonumber \\
&=&C (\cosh(\eta_0+\eta)+\sinh(\eta_0+\eta) \cos 2 \omega_0 t).  \label{12}
\end{eqnarray}
In particular, any solution without a breathing motion ($\eta_0=0$) will
be transformed to a breathing one and on the other side any breathing
motion can be transformed to $0$ by choosing $\eta=-\eta_0$.

As mentioned below Eq. (\ref{I}) all these properties are
a direct consequence of the underlying group structure and $\omega_0^2
C \cosh(\eta_0+\eta)$ can be identified with the energy of the system,
while the amplitude of the oscillations $ C \sinh(\eta_0+\eta)$ is
$2/\omega_0 \sqrt{\langle L_1\rangle^2+\langle L_2\rangle}$.  The
quantity 
$\tilde{C}=\langle H\rangle^2-(2
\omega_0)^2 (\langle L_1\rangle^2+\langle L_2\rangle^2)$ is conserved under a
group-transformation ($\cosh^2-\sinh^2=1$) and plays the role of the
Casimir invariant.  Generally, if $\tilde{C}$ is positive, one can
always find a group transformation, so that
$\langle L_1'\rangle^2=\langle L_2'\rangle^2=0$ and therefore
$I(t)'=\sqrt{\tilde{C}}/\omega_0^2=const.$ For $\tilde{C}<0$ this is
not possible and $I(t)$ describes the collapse of the system as
has been discussed in \cite{pitaevskii2}.


To conclude, we have shown that the existence of breathing oscillations
of $2D$ atoms in a harmonic trap is ensured by a hidden symmetry of the system.
If the atoms interact to a good approximation by a Fermi pseudo-potential,
i.e. by a local interaction, we expect well defined modes with a frequency of
exactly $2 \omega_0$. We identified the corresponding solutions of
\cite{kagan} with continuous representation of the underlying ``Lorentz'' group
SO(2,1).

The experimental conditions for the
realization of a 2D system in a magnetic trap look quite promising
in the case when most of the atoms are in the condensate. The system can be 
considered as 2D if $\omega _z \gg \omega _0$ is fulfilled. 
But our calculations are also valid for a system without condensation,
 e.g. for fermions. 
In this case the condition for two-dimensionality also demands
 that the effective temperature is less than the level
 separation in $z$-direction: $T\ll \hbar \omega _z$. 

On contrary one can also use a in $z$-direction prolonged  trap
 \cite{mewes} and
 excite  oscillations in the condensate with no 
$z$-dependence. 

 Important new possibilities
 are opened by a recent proposal \cite{pfau} to confine
 atoms in a 2D optical dipole trap. In such an experiment the appearance of
 sharp $2\omega _0$ frequencies would be  the first demonstration
 of the two dimensional nature of the system.
Experimentally it is also easy to excite the $2 \omega_0$- modes by a
 change of the external potential as described in \cite{kagan}.
Precision measurements of this response permit to check the assumptions of
the model.
Especially the validity of a local interaction, 
which was generally accepted so far, could be investigated.
It will be interesting to study deviations from such a description
 both experimentally and theoretically.

We acknowledge helpful discussions with Natan Andrei. L.P. likes to thank
the Institut f\"ur Theorie der Kondensierten Materie, Universit\"at
Karlsruhe, for hospitality and the Humboldt
Foundation for support.

\end{document}